# Not Only WEIRD but "Uncanny"? A Systematic Review of Diversity in Human–Robot Interaction Research

Katie Seaborn[1] · Giulia Barbareschi[2] · Shruti Chandra[3]

**Abstract**
Critical voices within and beyond the scientific community have pointed to a grave matter of concern regarding who is included in research and who is not. Subsequent investigations have revealed an extensive form of sampling bias across a broad range of disciplines that conduct human subjects research called "WEIRD": Western, Educated, Industrial, Rich, and Democratic. Recent work has indicated that this pattern exists within human–computer interaction (HCI) research, as well. How then does human–robot interaction (HRI) fare? And could there be other patterns of sampling bias at play, perhaps those especially relevant to this field of study? We conducted a systematic review of the premier ACM/IEEE International Conference on Human-Robot Interaction (2006–2022) to discover whether and how WEIRD HRI research is. Importantly, we expanded our purview to other factors of representation highlighted by critical work on inclusion and intersectionality as potentially underreported, overlooked, and even marginalized factors of human diversity. Findings from 827 studies across 749 papers confirm that participants in HRI research also tend to be drawn from WEIRD populations. Moreover, we find evidence of limited, obscured, and possible misrepresentation in participant sampling and reporting along key axes of diversity: sex and gender, race and ethnicity, age, sexuality and family configuration, disability, body type, ideology, and domain expertise. We discuss methodological and ethical implications for recruitment, analysis, and reporting, as well as the significance for HRI as a base of knowledge.

**Keywords** Intersectionality · Systematic review · WEIRD research · Generalizability · Diversity · Human–robot interaction

## 1 Introduction

People are diverse. While most of us may agree with this statement, we may also take it for granted. This is made clear if we consider a large-scale and ongoing pattern in human subjects research: the WEIRD sampling bias [1–4]. The acronym WEIRD, coined in 2010 by Henrich et al. [1], refers to the tendency for most human subjects research to sample people from Western, Educated, Industrialized, Rich, and Democratic societies. In a nutshell, the greater portion of the work on attitudinal and behavioural topics has drawn from undergraduate populations at Western universities [1]. When discovered, this pattern prompted a concerted effort to determine its extent within various domains and its significance for generalizing knowledge. Research generated from WEIRD populations has been, and continues to be, treated as *universal*, even though it captures only a slice of all human experience, and a *narrow* slice, at that [5]. Early research [1, 4] and more recent reviews [5–7] covering work in psychology, cognitive science, and economics has made it clear that universality is not a given, even while some knowledge tends to hold true across cultures and time. Moreover, recent work in human–computer interaction (HCI) has found that the same WEIRD patterns are at play [3]. As an adjacent, if not incumbent, field of study, HRI could be WEIRD, too. Indeed, the first objective of this research was to establish whether and to what extent this has been the case.

But does "WEIRD" capture the extent of sampling biases in HRI research? Critical scholarship, analyses, and whistleblowers within and outside of academia have raised awareness and called for action on broader matters of representation and diversity [7–14]. HCI researchers have pointed to limits in who is demarcated as "the user" in terms of envisioned designs, user groups, and participant pools [3, 9, 11, 14]. Others have pointed to biases present on the researcher and practitioner side: who is involved in research, who gets hired, whose ideas are selected for research grants and R&D, whose talent is sought out, how societal discrimination limits opportunities, education, and exposure, citation likelihood, and so on [14–16]. Researchers in AI have called out algorithmic bias of all kinds at all levels of production and study: assumptions in the rules making up the algorithms, unrepresentative training datasets, limited training protocols, and more [15, 17–19]. Buolamwini famously demonstrated how a facial identification algorithm used in computer vision failed to detect her face–she being Black–but had no problem detecting a white mask. She and collaborator Gebru later explicated their results and implications for the technical side in a landmark paper [20]. HRI researchers, especially feminist scholars and those invested in anti-racism work [21–24], have also echoed these concerns and produced artistic and academic work highlighting and tackling matters of identity and power with and through robots. Ladenheim and LaViers [25, 26], for instance, used a performance art approach to explore and provoke critical engagement with the feminine in the machine–how robots designed like human women do little more than reinforce stereotypes. Indeed, within and outside of HRI research, a range of critical voices have raised the alarm and provided evidence for oddities outside of the realm of WEIRD sampling. Gender and race have received the wealth of attention so far, but other factors may be overlooked. We thus turn to *intersectionality*, a legal model translated into an analytical framework that explains how power operates differently when multiple social and political identities are at play, leading to diverse experiences as well as different forms of discrimination [27, 28]. Indeed, the way that power operates through social structures and institutions, including academic fields, can be explained by a matrix of domination [29, 30]. These frameworks ask us to consider several more modes of identity beyond those represented in the WEIRD pattern, as well as demand that we address their intersections. Are there "uncanny" junctions among these factors within the participant populations invited to join HRI research projects? Our second objective is to find out.

✉ Katie Seaborn
seaborn.k.aa@m.titech.ac.jp

[1] Tokyo Institute of Technology, Tokyo, Japan

[2] Keio University, Yokohama, Japan

[3] The University of Waterloo, Waterloo, ON, Canada

In this systematic review of 749 peer-reviewed academic publications reporting on a total of 827 studies, we turn an intersectional lens on the question of sampling biases in HRI research. Drawing from previous work in adjacent fields, we first asked: (RQ1) Is HRI research WEIRD and to what extent? Then, drawing on extant critical theoretical frameworks, we also asked: (RQ2) Is HRI research limited in terms of diversity in other ways and to what extent? As a first step, we focused on describing the state of affairs through the representative case of a key venue–the premier ACM/IEEE International Conference on Human-Robot Interaction. We also aimed to highlight the presence and extent of these patterns over unearthing their impact, a significant effort given the amount of work to be covered, which we leave for future work. Our contributions are threefold. First, we contribute our participant diversity framework, which is grounded in critical scholarship within and beyond the field of HRI. Second, we offer evidence of WEIRD and "weirder" patterns in HRI research; or, in other words, sampling biases. Lastly, we map out the extent of each pattern across a large number of representative works published to a premier HRI venue. We urge our fellow researchers in HRI to take notice and reconsider the "who" for the "what" of knowledge creation.

## 2 Conceptual Framework of Diversity

The Uncanny Valley was first proposed by Professor Emeritus Masahiro Mori as a way of pinpointing and describing when and where robots approach, but not quite achieve, humanlikeness, thereby invoking feelings of strangeness and unease [31, 32]. In this work, we have considered the other half of the HRI equation: the human side. As our systematic review will show, there is something akin to uncanniness about the "who" in HRI research. Humanoid robots that do not quite approach true realism invoke a sense of disquiet; similarly, we should find it unsettling when the participant populations in our body of work are so narrow, that we know so little about the people we study, and that we underplay the importance of sampling. The WEIRD framework provides a solid starting point, but it is also limited, capturing only matters of culture, broadly framed, population education, national industrialization, economic output, and political orientation. Moreover, as a nation-level framework, it does not capture factors of identity, smaller-scale social groups, and other individual-level features that may play a role in human subjects research [33, 34].

We aim to address the limitations of the WEIRD framework and expand its purview by drawing on theories of *intersectionality*. Legal scholar Crenshaw coined the term "intersectionality" as a way to describe how exclusion and oppression intersect with multiple factors of identity in ways that are not necessarily additive, but often different [27]. Examples from her seminal work consider gender and race differences in the experiences of middle-to-upper class white women and Black women across different classes, i.e., how sexism, racism, and classism intersect. Collins [29] built upon this framework in her *matrix of domination*, illustrating how institutions and power structures create unique forms of discrimination and exclusion for African American women in contrast to white American women and African American men. This framework also accounts for how those with greater power are often unable to recognize or experience these forms of oppression, and may even benefit from them, i.e., privilege. The WEIRD framework represents an acknowledgement of how certain identities, characteristics, and social groups have been centred while others have been sidelined or treated as the same as those centred, i.e., recognizing intersectional privilege in human subjects research sampling. Yet, it does not cover all relevant factors, for participant populations in general and specifically for HRI research.

We thus developed a theory-driven multidimensional *framework of diversity* comprised of factors that can indicate whether and to what extent participant samples in HRI research are *diverse*. We relied on intersectionality theory [27] and the matrix of domination as a baseline [29]. These theories have only recently been situated within technology and design work. Notably, *intersectional design* has been conceptualized within HCI alongside traditions of human-centred practices and participatory design as *intersectional HCI* [35] and *intersectional computing* [36]. Taking a human-centred design perspective reduces the scope to the level of the user/s of the designed object, i.e., robots [30, 37, 38]. Centring the person this way also centres the factors originally identified by Crenshaw, Collins, and others, including gender, race, and class.[1] A human-centred design perspective also raises several more factors for consideration. For this, we drew on the person-level factors from the intersectional design framework represented in the design cards created by Jones and colleagues [37] and expanded upon it based on recent developments within HCI, HRI, and adjacent spaces. Our framework thus includes: sex alongside gender [37, 40]; ethnicity alongside race [14, 35, 37]; sexuality [37] as a factor of social identity linked to sex and gender, but also with implications for family configuration [37]; disability [37, 41, 42], which overlaps with but is distinct from the body [42, 43]; ideology beyond political affiliation and nation-level governance structures [44]; and domain expertise [45], particularly engineering, computer science, informatics, and related fields for HRI research.

As this is the first work to systematically appraise the field of HRI in this way, we focused on extracting and describing participant samples based on author reporting. Our framework, being based on intersectional theories and design frameworks, may also be used to assess intersectionality between two or more diversity factors. However, due to scope, we must leave such assessments to future work. We now turn to defining and justifying each factor with special attention paid to the nature of HRI research. We acknowledge that this selection is not exhaustive even while it represents most of the common elements in design-centred intersectional frameworks.

### 2.1 Sex and Gender

A wealth of research has called attention to the question of sex and gender [11, 24, 35, 46–51]. Often, the two are used interchangeably, although they are more usefully distinguished by sex as biology and physiology and gender as identity, expression, and social roles [52–54]. Moreover, a binary model prevails, especially in Western contexts [49, 54]. Yet, it has long been known that intersexual people with ambiguous sex characteristics exist [53]. Various cultures at different points in time have also acknowledged a range of genders beyond and within the masculine and feminine, such as Two-Spirit in Turtle Island cultures [55] and third genders in India [56]. People can be transgender, having a gender identity different from that assigned at birth according to apparent sex [57]. People can also be gender fluid, taking on characteristics generally associated with masculinity or femininity at the same time or different times. Others decentre or reject gender entirely, opting for gender neutral pronouns and referents such as "they." Cisgendered people are comfortable in the gender assigned to them at birth. While recent science and research on sex and gender has moved towards acknowledging and adapting to this diversity [40], take-up is slow and a bias towards cisgender, gender binary models remains prevalent. Additionally, a masculine bias in science generally and technology in particular has been well established [58]. When it comes to sampling biases in HRI, this may involve only recruiting men, typically on account of relying on undergraduate populations in engineering or computer science, which are primarily made up of men. Here, we seek to discover whether and how these biases map onto participant pools and reporting.

### 2.2 Race and Ethnicity

Race and ethnicity are two often intertwined but distinct social characteristics. Race refers to a way of categorizing people based on distinct physical features, while ethnicity is a broader concept, referring to a way of categorizing people according to shared cultural backgrounds and expressions that can be racial, geographic or national, religious or spiritual, and/or linguistic in origin [59]. Race and ethnicity have long been identified as axes through which social power can be explained, including within research and technology spaces [14, 35]. Notably, the modern academic world is anglocentric, oriented towards racial hierarchies and ethnic norms from British and American cultures [1, 2, 60, 61]. For example, English is the norm in academic publications [62] and may also be required for research even in countries where English is not the main language. The majority of research participants identify as white or Caucasian as well as being Western and English-speaking [63]. We attempt to extract race and/or ethnicity information about participants, paying particular attention to the possibility of anglocentrism as a likely pattern in sampling at the intersection of nationality, race and/or ethnicity, and language.

---

[1] We do not cover class in this work because it is an implicit and multi-dimensional variable, hard to operationalize and disentangle from other demographic factors, including age, education, race and ethnicity, family configuration, and national wealth. For example, many surveys offer a default "household income" option that could represent class (or not), but is dependent on the age of the respondent, the number of other members in the household, the number of dependents in the household, etc. [39].

## 2.3 Age

In most societies on earth at most times in the modern age, one can find people of various ages living, working, playing, and experiencing life. Yet, as WEIRD research has highlighted, much human subjects research involves undergraduate student populations, most of which are relatively young compared to the rest of the population. Moreover, some age groups are deemed "special populations," sequestered and given special focus in solitary studies, notably older adults and children. In this review, we map out whether and how this is the case in HRI.

## 2.4 Sexuality and Family Configuration

Sexuality refers to sexual and/or romantic orientations towards oneself and others, typically framed around the sex and/or gender of those involved [64]. Queer folk and sexual minorities are not heterosexual, or not exclusively. Family configurations are linked to sexual and romantic relationships, kinships, and other interdependencies among people, In most societies on earth, the dominant, centred, and/or expected sexuality is heterosexuality [65], with family configurations typically based upon cisgendered couples comprised of a man and a woman, i.e., heteronormativity [66]. The result of this pattern is the commonplace assumption that everyone is heterosexual and heterosexual norms in relationships apply to all relationships. When designing studies around families and carrying out our recruiting, we may assume, for instance, that a family unit is comprised of a mother, a father, and one or more children. Sexuality may also play a role in multi-user HRI contexts that consider intimacy or rely on assumptions about opposite-gender attraction, as well as work on sex robots. We consider whether and how diversity in relationships, sexualities, and family configurations is represented in HRI research.

## 2.5 Disability

Disability refers to ways in which people with impairments or different bodily configurations and neurocognitive patterns are encumbered or restricted in their interactions with the world due to limiting and/or restricting factors in the social and/or physical environment [67]. Most people will experience disability within their life. Disability can be visible or invisible, temporary or long-term, from birth or incidental, static or dynamic. Impairments can be external, i.e., limb impairment, or internal, e.g., kidney disease, or of the mind, i.e., cognitive impairments. Disability is often not an off/on state, but rather a continuum that can be context-dependent and change over time. For example, blind people may not be entirely without sight, able to see a range of shapes and movement even while qualifying for legal blindness status. Fundamentally, human bodies come in an ever-shifting variety of shapes and abilities. Nevertheless, most societies on earth tend to centre a certain range of bodily configurations and interaction capabilities, often to the exclusion of others, whether on purpose or incidentally [68]. Moreover, while disability rights activists and allies have raised attention to and fought for inclusion of those with visible, physical disabilities and certain learning disabilities, others continue to be sidelined. Many with hidden, internal or cognitive disabilities can "pass" as nondisabled, although this often involves great effort, personal expense, and constant vigilance [69, 70]. Activists and allies have rallied for awareness and support around these forms of disabilities, advocating for recognition of the neurodiversity of people as well as implicit assumptions of neurotypicality.

Research can be disabling, such as when researchers do not consider how a person in a wheelchair can enter a building to join a study, or when lab assistants do not provide clear directions to people with dyslexia. Even when the specific goal of the research is to create technology that empowers people with disabilities, incorrect assumptions around capabilities, needs, and preferences that originate in designers' and researchers' lack of lived disability experience can lead to negative results [16]. Since the ratification of the UNCRPD (United Nation Conventions on the Rights of People with Disabilities) in 2006,[2] the disabled community has advocated for "Nothing About Us Without Us": full inclusion in design, research, and other forms of work on disability. In recent years, the realization of the importance of this commitment in accessibility research has greatly matured [41]. Ultimately, people with disabilities represent the largest minority of individuals in most countries worldwide [71]. When developing inclusive HRI technology, we need to ensure representation of people with disabilities in HRI research.

## 2.6 The Body

Interacting with robots often involves physical presence, if not the use of one's body. Height may be a key index in designing appropriate interactions. For instance, it would be difficult for a tall person to shake the hand of a small robot, if both are standing on a flat surface. Other features, such as hand and finger size, handedness (right, left, ambidextrous), gait, and even hairstyle are known to have implications for technology and potentially robots [72]. In *Black Klansman* (2018), Rob Stallworth recounts how he was given a helmet to wear as part of his police officer uniform, which had not been designed to consider Black hair and especially afros. Given the resurgence of virtual reality, especially headwear, and the field's emerging intersections with robotics [73], we should consider whether and how different bodies and bodily expressions are accommodated. More subtly, recent work on fatphobia and fat exclusion [74] requires us to take a critical look at our participant pools: could there be unconscious bias in recruitment at play? Certain body features, including weight, height, and proportions [75], have a long history of marginalization in human factors, especially in industrial design.

---

[2] https://www.un.org/development/desa/disabilities/convention-on-the-rights-of-persons-with-disabilities.html.

Critical scholars active in the field of HCI have highlighted how the systematic exclusion of bodies deemed non-normative, often with the implication of deviance in need of correction, leads to the development of technologies that oppress rather than empower individuals, who are often already marginalized [76, 77]. For instance, an influential critical race paper by Ogbonnaya-Ogburu et al. [14] highlights how uncritically conducting research that features racialized bodies merely reinforces existing dynamics of privilege and produces artificially unrepresentative results. Recent reviews by Gerling and Spiel [78] and Spiel [42] have critically analysed the existing literature on embodied interactions and virtual reality technology, respectively. They have identified how implicit assumptions about bodies and embodiment have steered the development of these technologies towards reinforcing an existing dichotomy of normative, "ideal" bodies versus non-normative, "deviant bodies." This needs to be corrected, or more simply eliminated. To the best of our knowledge, implications about how the size, shape, and other physical characteristics of human bodies affect interactions with robots have not been explored to date. In this review, we seek to find out the extent to which this history has played out in HRI research.

### 2.7 Ideology

Ideology is an umbrella term for "a system of ideas and ideals, especially one which forms the basis of economic or political theory and policy" (Oxford Languages). Ideology is formed at different social levels, in families and communities, in societies and cultures [79, 80]. Social institutions, such as political parties, organized religions, and even online communities can play a role in shaping one's ideological stance [81]. Technology has been disruptive in this regard, such as with algorithmic bias on social media and digital radicalization [17, 82]. Often, there is a combination of ideological forces at play. For example, one can identify as a Christian, but one's values and beliefs may vary from other Christians, especially at other times in history and across cultures. Ideology has implications for values, beliefs, attitudes, behaviour, decision-making … essentially, how we make sense of and interact with the world. Roboticists have started to explore ideological frames directly, such as with spiritual robots [83], robots that assist in religious practices [84], and robots that question beliefs and provoke critical thinking [85]. More subtly, research constructs that are not directly coded as "ideological" may be implicated by one's ideology. For example, Indigenous cultures may not draw the same distinctions between humans and robots that Western frameworks do [86]. Do researchers take on an all-knowing stance when it comes to participants, their ideologies, and how these ideologies affect the variables under study? We seek to find out whether and how ideology has been considered.

### 2.8 Domain Expertise

WEIRD research in HCI has shown that participants are often sampled from the universities at which the studies are being run, if not the specific departments to which the researchers are affiliated [3]. We expect this pattern to take place within HRI research, as well. Specifically, we expect to find that the greater portion of participants have computer-oriented backgrounds. Students in engineering programs, professional engineers and programmers, roboticists, technicians, industry professionals working in technology spaces … there may be a breadth of positions, but all will fall within computer-oriented domains. This has implications for knowledge-building and practice. The knowledge implications have been well-mapped by WEIRD research [1, 2]. But HRI often seeks to develop robots that will someday be embedded in people's lives. Having specialist knowledge means knowing, to some degree, the capabilities and limits of robots, especially when breakdowns may occur and why. This knowledge affects initial and ongoing acceptance, trust and reliance, extent of use, and so on [87]. In short, people who are domain experts may have different attitudes and behaviours towards robots compared to the general public. We should map out how frequently these experts take on the role of participant.

## 3 Methods

We conducted a systematic review and content analysis on the nature of the populations sampled in works published to the ACM/IEEE International Conference on Human-Robot Interaction (HRI) from its inception (2006) to the present year (2022). We chose this venue because it is considered the premier conference on HRI research as well as one of the highest cited in social robotics generally, with an h-index of 50 and an h-median of 71 in 2022.[3] Our review was guided by the PRISMA approach [88], modified in line with the standards for our discipline, i.e., most venues do not require structured abstracts or PICO/S. Our PRISMA flow chart is shown in Fig. 1. Our protocol was registered in advance of data collection on May 5, 2022.[4]

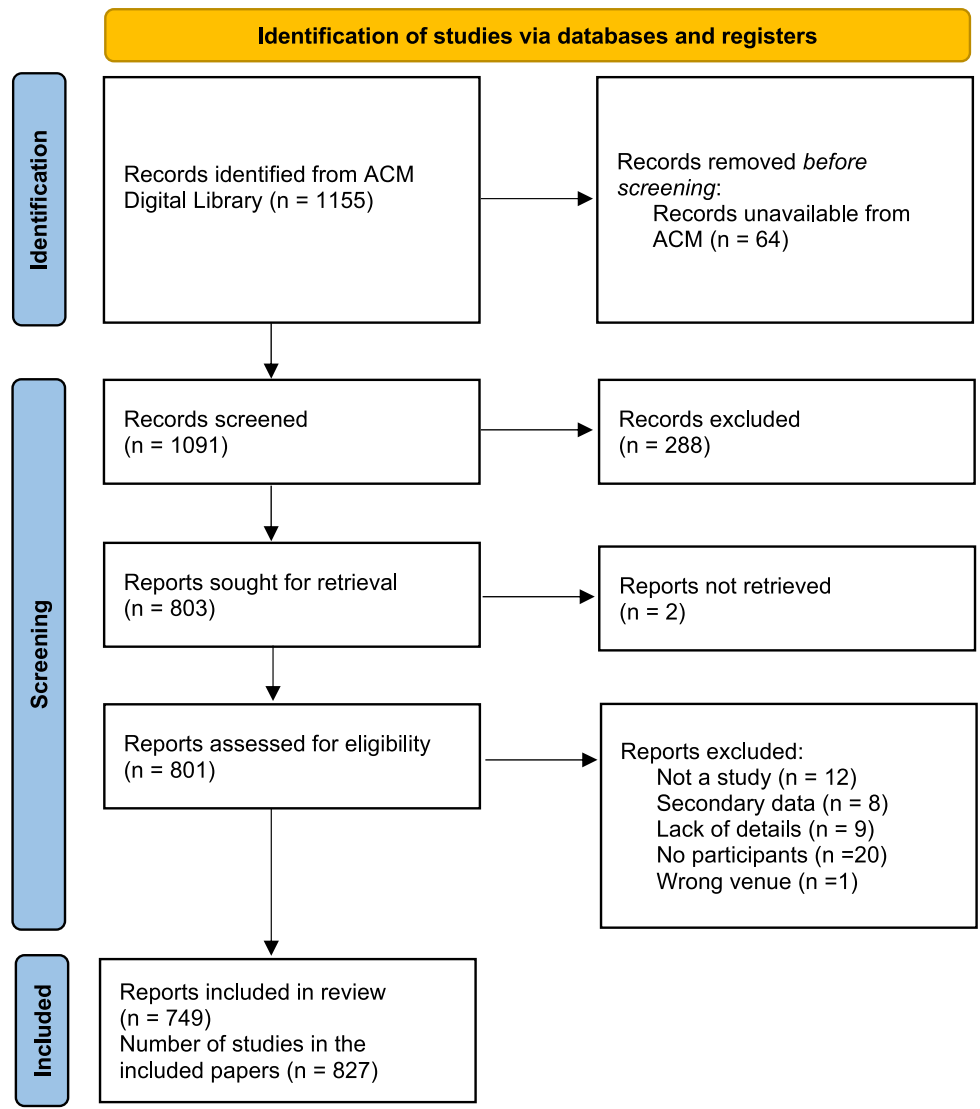

**Fig. 1** PRISMA flow chart

### 3.1 Eligibility Criteria

Papers were included if published as a full or short paper in the HRI proceedings and included human subjects research with at least one participant. Papers were excluded if inaccessible, pilot studies with insufficient detail to establish the inclusion criteria, or a preprint.

### 3.2 Information Sources and Search Strategies

We used the ACM Digital Library (DL) to search the HRI conference proceedings. We constructed the following meta query: *Human ("human subject*" OR participant*) AND Robot AND (robot*) Interaction AND ("human–robot interaction*" OR hri)*. Since the ACM DL does not allow the selection of all conference proceedings by name, we modified the query to restrict to the HRI conference proceedings.

---

[3] https://scholar.google.com/citations?view_op=top_venues&hl=en&vq=eng_robotics.

[4] https://osf.io/jtnqz.

The full query was: *[Publication Title: "conference on human robot interaction"] AND [[Abstract: "human subject*"] OR [Abstract: participan* user*]] AND [Abstract: robot*] AND NOT [Abstract: survey] AND NOT [Abstract: "literature review"]*. The query was run on May 6, 2022, resulting in an initial 1155 items. Even so, due to an apparent glitch in the ACM DL system,[5] only 1091 were available to be downloaded.

### 3.3 Selection of Data, Data Collection Process, and Data Items

The first author downloaded the query results, exporting the metadata into Zotero. Items that were not papers, such as conference proceedings outlines, were removed at this stage. The first author then screened all of the 1091 items alone based on the abstract. The second author then checked the excluded items. The full text was checked when disagreements occurred. The three authors then divided the resulting 801 items amongst themselves for full text screening. As before, items marked for exclusion were double-checked by another author, and disagreements were resolved through discussion. Data items extracted were participant details and descriptions of instruments or measures, if relevant to the factors under study, as well as relative paper size (short, up to four pages, or long, over four pages).

### 3.4 Data Analysis

We classified the extracted data in line with Henrich et al. [1] and Linxen et al. [3]. We generated descriptive statistics for all variables, including, where appropriate, counts, percentages or ratios, means, medians, standard deviation, interquartile range. When the data was not available within the paper or through tertiary sources, we marked it as such. We then generated ratios (Sect. 3.4.1) to represent how WEIRD each sample appeared to be, following Linxen et al. [3]. Next, we evaluated each WEIRD variable individually (Sects. 3.4.2–3.4.6). We then turned to analyzing our diversity factors (Sect. 3.4.7; refer to Sect. 2). Next, we analyzed the WEIRD and diversity factors together, aiming for triangulation and consensus (Sect. 3.4.8). Finally, we considered the influence of page length on all results (Sect. 3.4.9). We describe how we conducted each of these analyses in detail next.

### 3.4.1 Overall WEIRDness

We classified the overall "WEIRDness" of recruited populations across the corpus of papers. Unlike Henrich et al. [1], we differentiated between the reported location of the study and the self-reported nationality of participants. We assumed that most studies would be conducted at the university, which are often multicultural environments that include people of different nationalities [89]. Yet, while extracting the data, we noticed that authors were significantly more likely to report the location of the study rather than the nationality of participants, which was reported in less than 10% of the studies. In light of this, we conducted two analyses for nationality. We first conducted analyses on the reported *location of the study*. We then analyzed the 10% of papers that provided the *self-reported nationality* of participants. We used the same formula as Linxen et al. [3] to normalize the number of participants in a study ($\varphi$) by their country's population using data from the World Bank.[6] We thus generated a participant ratio ($\psi$) using Linxen et al. [3]'s formula:

$$\psi = \frac{\# \, of \, \varphi(country) \cdot population(world)}{\# \, of \, \varphi(total) \cdot population(country)}$$

This formula generates a ratio value. A value of 1 means that the number of participants or participant samples are proportional to the nation's overall population. If the value is greater than 1, the nation is overrepresented. If it is under 1, then the nation is under-represented. Based on this ratio and the factor-specific quantitative results, we made a qualitative team-based summary judgment about the representation for each WEIRD and diversity factor.

### 3.4.2 Western

We classified countries as Western and non-Western according to the criteria of Henrich, Heine, and Norenzayan [1] criteria. Specifically, Western countries were deemed to be those located northwest of or in Europe (the United Kingdom, France, Germany, etc.), and Western colonized nations, including the United States, Canada, New Zealand, and Australia. Given the state of reporting, we decided to distinguish location of sampling and reported nationality of participants. For example, if participants were recruited on the university campus, we did not assume that they were of the nationality associated with the location of the university. If nationality was not reported, we marked these participants' nationalities as not available. Additionally, some reported participants' language ability, e.g., Korean speakers, which could be taken as an implicit marker of nationality. However, because people can learn multiple languages or be expected to use the language/s associated with the location of the study, we did not assume that language ability indicated national origin and so marked these cases as not available.

---

[5] Although the ACM Digital Library returned a count of 1155, we could not find 1155 items when traversing the pages of results: we could only find 1091. We are not sure if the count was incorrect or if a selection of results was not made available or skipped for some unknown reason.

[6] https://data.worldbank.org/indicator/SP.POP.TOTL.

### 3.4.3 Educated

We classified education data according to the 2011 International Standard Classification of Education (ISCED 2011), using Eurostat's aggregated levels.[7] Specifically, low education used Levels 0–2 (early, primary, and lower secondary), middle education used Levels 3–4 (upper secondary and post-secondary), and high education used Levels 5–8 (tertiary, bachelor's, master's, and doctoral). We also counted "no education."

### 3.4.4 Industrialized

Since industrialization is a country-level factor, we followed Linxen et al. [3] in using the gross domestic product per capita (GDP) adjusted with purchasing power parities (PPP) to account for differences across countries unrelated to economic industrialization.

### 3.4.5 Rich

As in Linxen et al. [3], we used the gross national income per capita (GNI) adjusted with PPP. The GNI captures the flow of wealth within and outside of a country and acts as an indicator of living standards for the average person in that country.

### 3.4.6 Democratic

Like Linxen et al. [3], we used the political rights rating as a measure of each country's democratic standing according to Freedom House.[8] Political rights covers an array of governmental characteristics and activities at a societal level, including voting, individual participation, political pluralism, and so on.

### 3.4.7 Diversity Factors

We relied on manual extractions about participants in the paper to derive statistics on the diversity factors outlined in our conceptual framework. Most of these factors were categorized nominally and analyzed by frequency of appearance within our data set because the corresponding national statistics did not exist. Specifically, size characteristics, many forms of disability, neurotypicality and neurodiversity, and certain ideological frames were not typically available at a national scale from reliable sources at the time of data analysis. Even when such data existed for certain variables, e.g., sexuality, ideology and religion, and certain disabilities, we found that most of these were not described in the papers. For domain expertise, we distinguished between computer familiarity, robot familiarity, students in computer science and/or engineering (CS/eng), and experts in robotics. We note these when reporting our results. All categorizations were double-checked by at least one other researcher to ensure rigour.

Sex and gender, age, and race and ethnicity, however, were treated in a similar fashion as the WEIRD variables. For gender, we used statistics from the World Bank[9] to generate ratios. Importantly, most data sets and research reports do not operationalize or distinguish sex and gender and assume a binary model. However, change is on the horizon; for example, Canada is one of the first countries in the world to distinguish gender and sex and provide a diverse range of gender identity options, including transgender and Two-Spirit.[10] For the time being, we acknowledge this limitation about our "gender ratio" data.

Likewise, we used statistics on age from the World Bank to generate ratios indicating the relative youthfulness of the samples. We used the mean and age ranges reported in each paper. We used two metrics to determine the cut-off point for age. One was the UN's classification of "youth" as between ages 15 and 24[11] and the WHO's definition of old age as 60 and above.[12] The other was the "emerging adulthood" classification of up to age 30 [90]. We recognize that this is a rough measure. However, many papers relied on nominal age categories or age ranges, which could not be extrapolated for comparative analysis. Finally, we used statistics from the World Bank on race/ethnicity to generate whiteness ratios. Ratio variables were treated the same way as the WEIRD variables, i.e., t-tests, correlations.

### 3.4.8 Length of Paper

The HRI conference offers two paper formats: short ($\leq 4$ pages) and long ($\geq 5$ pages), not including references. We realized that length of paper, i.e., space available to report details, can limit reporting and may thus act as a confounding factor. We therefore identified and calculated the relative influence of short and long papers on reporting. We did this by extracting the number of pages, excluding pages only used for references, based on the HRI conference guidelines and rules for page lengths and ranges. This produced two groups: "short" and "long" groups. We then re-conducting the analyses above for each group and compared the results.

---

[7] https://ec.europa.eu/eurostat/statistics-explained/index.php?title=International_Standard_Classification_of_Education_(ISCED)#Implementation_of_ISCED_2011_.28levels_of_education.29.

[8] https://freedomhouse.org/countries/freedom-world/scores.

[9] https://databank.worldbank.org/source/gender-statistics.

[10] https://www150.statcan.gc.ca/n1/daily-quotidien/220427/dq220427b-eng.htm.

[11] https://www.un.org/en/global-issues/youth.

[12] https://www.who.int/health-topics/ageing.

### 3.4.9 Archetypes: WEIRD and Diverse

We considered whether the WEIRD and diversity factors pointed to archetypes of participants, both included and excluded. We did this by sorting and splitting the data, calculating descriptive statistics, and thematically summarizing the quantitative results for individual factors. Possible intersections were informed by critical theories, global statistics (e.g., WHO reports), assertions and gaps in the included papers, and colloquial trends known to the authors.

## 4 Results

From an initial 1155 papers, 749 papers (423 short and 326 long) and 827 studies were included. The full data set is on OSF.[13] We now present the results by order of data analysis (Sect. 3.4), starting with an overview for WEIRD (Table 1) and diversity (Table 2) factors.

### 4.1 Study Locations and Participant Nationalities

259 studies (30.9%) reported on study location while 74 studies (8.8%) reported on the (self-reported) nationality of participants (e.g., *"We recruited 52 Korean people"*). Of these, 68 included an explicit count of the number of participants recruited according to their nationality (e.g., *"We recruited 15 participants from Japan"*). The remaining six provided participants' self-reported nationality as a list, without including the specific numbers of participants for each nationality (e.g., *"We recruited participants from the UK and the US"*). We extracted 272 country locations (Fig. 2). Several studies reported multiple locations. Overall, study locations totalled 31 countries, 13 of which only appeared once. 12 countries were mentioned as study location at least 5 times (112 USA, 42 Japan, 18 Germany, 10 Denmark, 10 Netherlands, 10 Sweden, 9 UK, 9 South Korea, 5 Austria, 5 Canada, 5 France, 5 Italy). Countries for which less than 5 studies were reported (Portugal, Lebanon, Kazakhstan, Singapore, China, Ireland, Qatar, New Zealand, India, Belgium, Greece, Turkey, Finland, Switzerland, United Arab Emirates, Mexico, Panama) were clustered under the label "Other Countries."

For participant nationality, 34 countries were reported for 9977 participants in 68 studies (Fig. 3). 26 countries were listed as a nationality for more than one participant. 12 countries were listed as a nationality for at least 100 participants, representing about 10% of the total number of participants for which nationality details were provided. The most reported countries were: 6363 for the USA, 1138 for Japan, 273 for Austria, 246 for Germany, 228 for China, 213 for Ireland, 208 for Italy, 171 for India, 164 for Sweden, 109 for South Korea, 103 for South Africa, and 100 for Denmark. Countries for which less than 100 participants were reported include Canada, New Zealand, Australia, UK, Spain, Belgium, Netherlands, France, Lithuania, Slovakia, Russia, Ukraine, Portugal, Poland, Cyprus, Mexico, Brazil, Peru, Egypt, Nigeria, Kazakhstan, and Israel.

### 4.2 Overall WEIRDness

We now present the ratios that represent an estimate of the extent to which each country was over- or under- represented (Fig. 4). In total, 17 countries were "over-represented" ($\Psi > 1$) and 17 were under-represented ($\Psi < 1$). The ten most over-represented countries by order of magnitude were: Ireland, Austria, USA, Denmark, Sweden, Japan, Israel, New Zealand, Netherlands, and Italy. Figure 4 shows a world map of the countries from which participants were recruited and not recruited. This visualization shows the relative over-representation of Western nations and under-representation of non-Western nations. Note that the extremely low numbers of participants for which nationalities were reported suggests that we should take caution when interpreting the ratios for over- and under-representation. Extremely small variations in the number of participants can significantly affect the $\Psi$ ratio.

Table 3 presents the number of studies, participants, and representativeness ratio for the 10 countries most frequently reported as the study location or as the nationality of participants. As a result of overlaps between the two sets, 16 countries are included in the table.

### 4.3 WEIRD Factors

We now present the results for each WEIRD factor individually and in detail.

#### 4.3.1 Western

Of the 31 countries listed as study locations, 18 were classified as Western and 13 were not. Although this might suggest an overall balance between Western and non-Western countries, a review of the number of studies conducted in each location shows that this was not the case. Of the 272 studies for which a study location was provided, 203 (74.6%) were conducted in Western countries, and 69 (25.4%) were conducted elsewhere. Similar patterns were also observed for participant nationalities. Overall, 20 (of 34) countries (58.8%) were Western and the remaining 14 (42.2%) were not. Considering the proportional representation of participants, 7902 individuals (79.2%) were reported to be nationals of Western countries, and only 2075 (20.8%) were reported to have non-Western nationalities.

---

[13] https://osf.io/thdvk/.

Table 1 Overview of WEIRD factors. This table represents the WEIRD status of all studies. Reported and Unreported rows highlight the studies from which an assessment of each factor could be made (e.g., Western or not). Representation is a qualitative summary of the quantitative results

|  | Western | Education level | Industrialization | Rich status | Democratic status |
| --- | --- | --- | --- | --- | --- |
| Reported | 259 studies | 270 studies | 259 studies | 259 studies | 259 studies |
| Unreported | 568 studies | 557 studies | 568 studies | 568 studies | 568 studies |
| Representation | Over | Over | Greatly over | Over | Over |
| Factor | Western | Highly educated | GDP PPP | GNI PPP | Free |
| Count (by study location) | 203/69 (western/non-western) | 239/46 (high/low + mid + no education) | 268/4 (GDP PPP > 18,724/GDP PPP < 18,724) | 266/6 (GNI PPP > 20,000/GNI PPP ≤ 20, 000) | 261/11 (free/partial free + non-free) |
| Ratio (calculated from the count) | 2.9 | 5.2 | 14.5 | 6.8 | 2.9 |

### 4.3.2 Educated

Only 270 studies (32.6%) unambiguously reported on education level, with 557 (67.4%) not reporting or not reporting with enough detail to determine the education level of all participants. Overall, most studies reported that participants were highly educated (239 studies or 28.9%), with a few studies reporting on participants having middle (24 or 2.9%) or low education (21 or 2.5%), and one study reporting on participants with no education. The highest educated participants in the West were located in the US (48 studies) and Germany (12 studies), while those from the East were located in Japan (16) and South Korea (6). A t-test found no significant difference between Western nations (83.7%) and non-Western nations (78.1%) in terms of representation of participants with a high education level, $X^2(2, N = 826) = 0.436, p = 0.509$. Most studies did not report exact counts, ratios, or percentages. For example, Gurung et al. [92] gathered participants through the university's "communication channels," but it is not clear whether and to what extent those using these channels were educated, at that university or elsewhere. Similarly, Xu and Dudek [93] reported that 86% of their participant pool were graduate students, but did not report on the education status of the other 14%. Due to reporting issues, our results may be unrepresentative of the actual population.

### 4.3.3 Industrialised and Rich

Almost all countries (27 of 31) listed as study locations were classified as high income, according to GNI PPP per capita. The only exceptions were India (GNI PPP 7220 USD), Lebanon (GNI PPP 10,360 USD), China (GNI PPP 19,170 USD), and Mexico (GNI PPP 19,540 USD). Even so, we should note that universities and other institutions may be located in areas of higher income and/or be places where people of higher income brackets live, work, and learn. Similarly, 29 (of 31) countries listed as study locations had a higher GDP PPP than the world average of 18,724 USD. India (7333 USD) and Lebanon (10,691 USD) were the only two countries listed as study locations with relatively lower GDP PPP rates, indicating a lower degree of industrialisation. In total, only six (2.2%) studies were carried out in non-high-income countries and four (1.5%) in non-industrialised countries.

When looking at countries listed in relation to participant nationalities, 25 were classified as high income and nine were not (Brazil, China, Egypt, India, Mexico, Nigeria, Peru, South Africa, Ukraine). In total, 27 countries had a higher GDP PPP than the world average of 18,724 USD. Brazil, Egypt, India, Nigeria, Peru, South Africa, and Ukraine were the seven countries with a reported GDP PPP lower than the world's average. Only 543 (5.4%) participants were listed as from non-high-income nations and 288 (2.9%) were listed as from non-industrialised nations.

### 4.3.4 Democratic

Most of the countries listed as study locations were considered "Free" according to the Global freedom score provided by Freedom House. In total 23 countries were classified as "Free", three as "Partially Free" (India, Singapore, and Lebanon), and five as "Not Free" (China, Kazakhstan, Qatar, Turkey, United Arab Emirates). Average score across all countries was as high as 75.7. When looking at the number of studies carried out in Free, Partially Free, or Not Free countries, the frequencies were 261(96%), 5 (1.8%), 6 (2.2%) respectively.

**Table 2** Overview of diversity factors. Representation is a qualitative summary of the quantitative results

| | The body | Ideology | Domain expertise | Race and ethnicity | Gender and sex | Disability | Sexuality and family configuration | Age |
|---|---|---|---|---|---|---|---|---|
| Reported | 102 | 95 | 323 | 206 | 156 | 150 | 14 | 349 |
| Unreported | 725 | 728 | 504 | 621 | 671 | 677 | 814 | 478 |
| Representation | Slightly over | Uneven | Over | Slightly over | Even* | Over | n/a | Under, over |
| Coding Variable | Implicit (Im) Explicit (Ex) | Morals (M) Identity (I) Beliefs (B) Religion (R) Politics (P) Law (L) | Computer familiarity (CF) Robot familiarity (RF) Students in CS/Eng. (SE) Experts who create research (Ex) | White English Anglo centric | Men Women Other Trans | Assumptions causing exclusion (Ae) Purposeful exclusion (Pe) | Heteronormativity | Youth (Y1) Young Adult (Y2) Younger (Y3) |
| Factor | Implicit > explicit (clarity) | Morals, identity, and beliefs > religion, politics, and law | Computer and robot familiarity > students and experts | Any combination | Men > women (*gender binary) | Exclusions by assumption > purposeful exclusion | Normative (N) > Obscure (O) | Youth, young adult and younger |
| Count (reported/unreported) | 102/725 | 95/728 | 323/504 | 206/621 | 503/324 (either M, W, O, T), 14/813 | 150/677 | 13/814 | ≤24: 13, ≤30: 57, ≥60: 19 |
| Ratio | 1.4 (Im/Ex) | 0.5 (M/all, 0.4 (I/all), 0.4 (B/all) > 0.1 (L/all), 0.03 (P/All), 0.1 (R/all) | 0.6 (CF/RF + SE + Ex), 0.6 > 0.2 (SE/CF + RF + Ex), 0.03 (Ex/CF + RF + SE) | 1.2 (W or E or A/(Non-W or Non-E or Non-A) | 1.1(M/W), n/a | 5.3 (Ae/Pe) | 0.17 (N/O) | 0.5 (Y1/Y2 + Y3), 1.9 (Y2/Y1 + Y3) |

*Gender/sex ratio was calculated on the basis of the binary assumption of participants being categorized as men and women

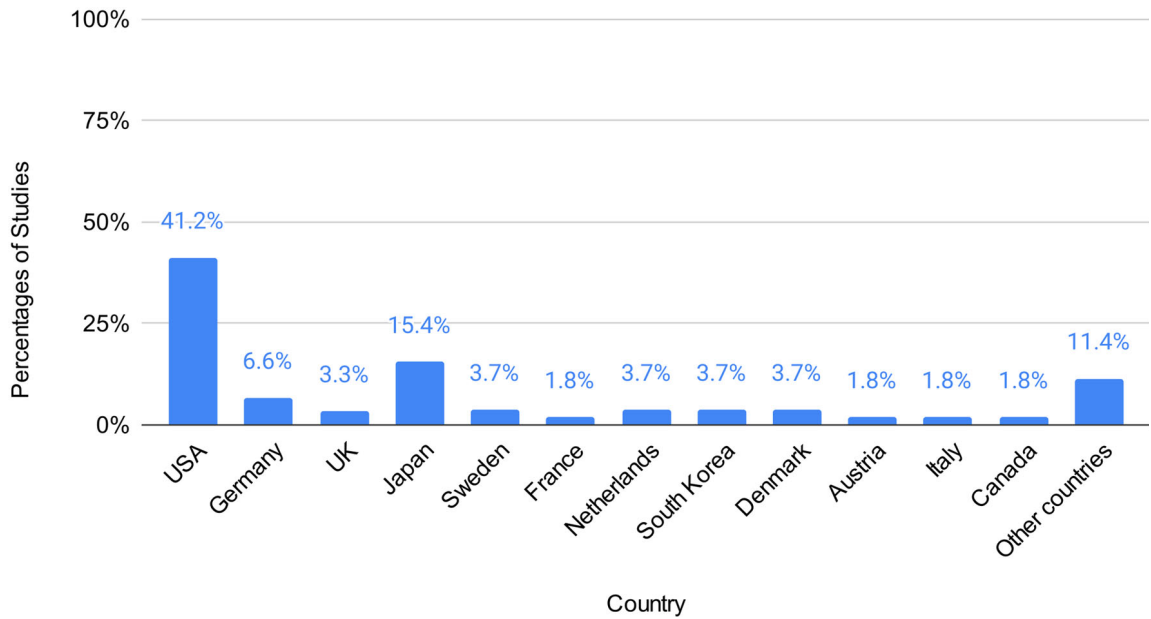

**Fig. 2** Countries reported as study location and frequencies. Other countries (frequency of < 5) include Portugal, Lebanon, Kazakhstan, Singapore, China, Ireland, Qatar, New Zealand, India, Belgium, Greece, Turkey, Finland, Switzerland, United Arab Emirates, Mexico, and Panama

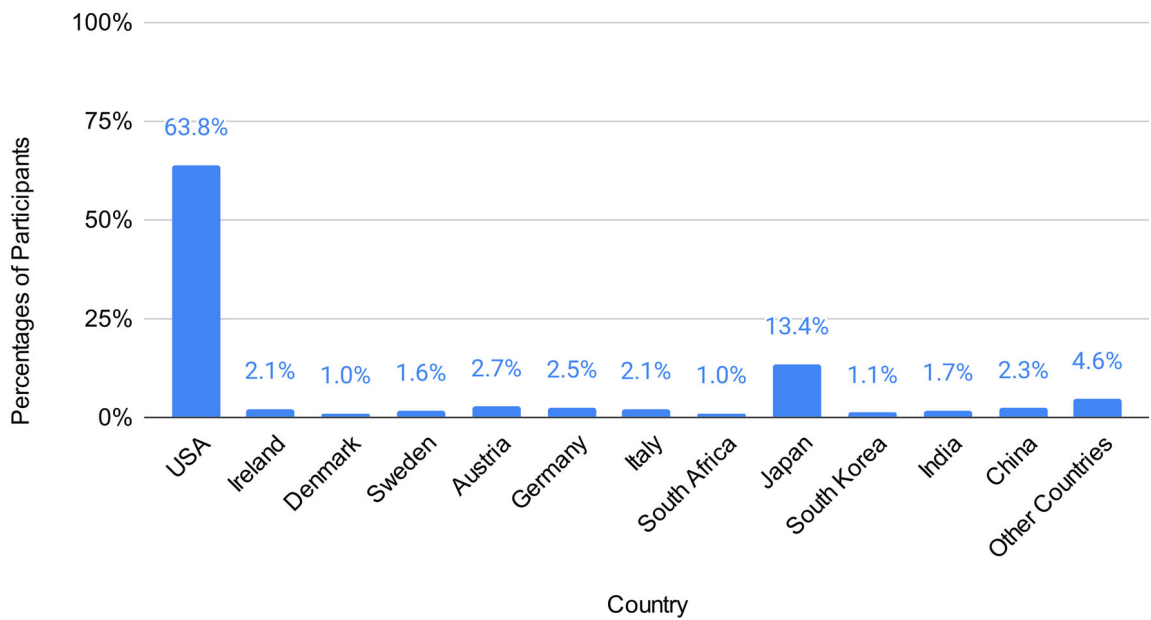

**Fig. 3** Participant nationalities and frequencies. Other countries (frequency < 100) include Canada, New Zealand, Australia, UK, Spain, Belgium, Netherlands, France, Lithuania, Slovakia, Russia, Ukraine, Portugal, Poland, Cyprus, Mexico, Brazil, Peru, Egypt, Nigeria, Kazakhstan, and Israel

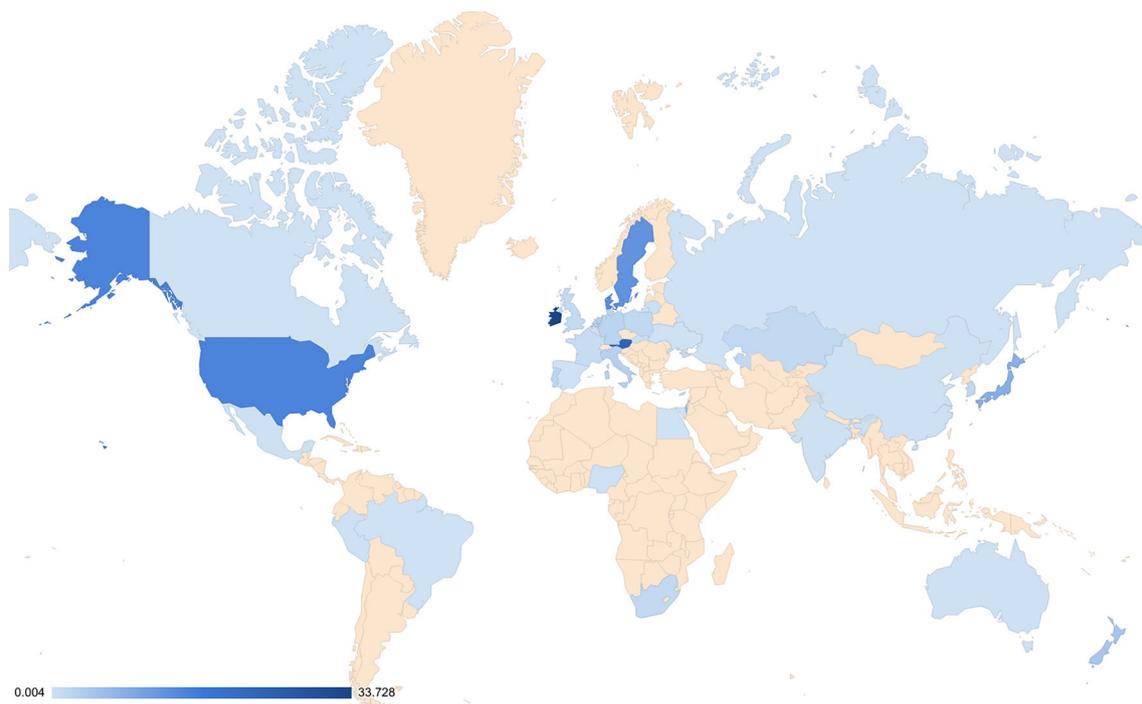

**Fig. 4** World map indicating the relative degree of over-representation (dark blue) or under-representation (light blue). Countries not mentioned in any study are highlighted (light orange)

When looking at the countries reported in relation to participants' nationalities, the pattern appeared to be similar. Twenty-seven out of 34 countries were classified as "Free", three as "Partially Free" (India, Nigeria, Ukraine[14]), and four as "Not Free" (China, Egypt, Kazakhstan, Russia). The average freedom score was 77.6. Finally, 9538 participants (95.6%) reported their nationality to be associated with a "Free" country, 173 (1.7%) with a "Partially Free" country, and 266 (2.7%) with a "Not Free" country.

### 4.4 Diversity Factors

We now turn to presenting the results for each factor in our diversity framework.

#### 4.4.1 Sex and Gender

A total of 787 out of 827 (95.2%) of studies reported the sex and/or gender of 55,032 participants. There were 15,523 men (28.2%, M = 31.2, MD = 17), 14,310 women (26%, M = 28.7, MD = 15), two trans* people, and 332 (0.6%, M = 6.3, MD = 1) unreported. A paired t-test did not find a significant difference between the numbers of men and women, $t(29,832) = 0.862$, $p = 0.39$, 95% CI [− 1.9623,

1.9623]). Sexuality was only explicitly reported in one study [94].

A gender-expansive approach was taken in 52 (10.3%) studies that reported on sex/gender. This involved including non-binary options or allowing for non-reporting explicitly (17, 3.4%) as well as being considerate of participants' privacy (35, 7%). For example, Steinhaeusser and Lugrin [95] reported on a participant who "self-reported as diverse gender," which is gender-expansive as well as considerate of privacy, with no specific gender details reported. Nevertheless, 106 studies (21.1%) relied on a gender binary approach. Several studies (32, 6.4%) wrote about aiming for, achieving, or failing to achieve a "balance" in participant numbers by gender, implying between women and men. 74 studies (14.7%) reported either men or women counts only, with 61 (12.1%) reporting only women counts and 13 (2.6%) reporting only men counts. The choice of reporting either men or women counts alone was significantly different, favouring the reporting of women counts, $X^2(1, 74) = 31.135$, $p < 0.001$. It is unclear why. Some authors commented on disproportionate counts between men and women participants. For example, Jensen et al. [96] wrote about an "uneven distribution of gender," while Karreman et al. [97] reported on male-male and male–female pairs, but no female-female pairs, on account of being unable to recruit equivalent numbers of female participants. Yet, there was virtually no

---
[14] Note that the freedom scores for both Russia and Ukraine were calculated prior to the start of the ongoing war.

**Table 3** Nations most reported as the study location or participant nationality

| Country | N-as study location | N-participants with nationality | Ψ |
| --- | --- | --- | --- |
| USA | 112 | 6363 | 14.95 |
| Japan | 42 | 1338 | 8.15 |
| Germany | 18 | 246 | 2.28 |
| Sweden | 10 | 164 | 12.63 |
| Denmark | 10 | 100 | 13.39 |
| Netherlands | 10 | 63 | 2.85 |
| South Korea | 10 | 109 | 1.65 |
| UK | 9 | 54 | 1.12 |
| France | 5 | 75 | 0.89 |
| Austria | 5 | 273 | 23.57 |
| Italy | 5 | 208 | 2.66 |
| Canada | 5 | 3 | 0.06 |
| China | 1 | 228 | 0.12 |
| Ireland | 3 | 213 | 33.72 |
| India | 1 | 171 | 0.1 |
| South Africa | 0* | 103 | 1.36 |

*A number of participants from one study [91] were reported as being from South Africa, but the location of the study was not explicitly reported

difference in the number of men and women within each study overall.

We also found a few cases of unconscious gendering (4, 0.8%) and cissexism (5, 1.1%). However, we recognize that not all authors may be fluent in English, the required language for publication, and may also rely on computer translations, which are notoriously sexist [98]. Rossi et al. [99], for example, created a "barman" robot, an example of unconscious gendering based on role, where "barkeep" would be the gender neutral equivalent. More subtly, Begum et al. [100] justified their exclusion of children with autism who were not boys based on rates of autism by sex/gender, which have been called into question [101]. More to the point, zero is not equivalent to "less than." In an example of cissexism, Choi et al. [102] decided to recruit from a women's college because "typically women are in charge of cleaning their houses." Similarly, Ise and Iio [103] decided to exclude women because of the gender and age combinations they wished to recruit but also because in their pilot tests they felt that "females tend[ed] to have more utterances." Von der Pütten et al. [104] prescribed the sex/gender of participants based on video. In contrast, Suomalainen et al. [105] wrote a frank discussion of gender in relation to VR sickness, with evidence for and against, ultimately deciding to recruit men and women while consciously excluding those who "preferred not to report their gender" towards this goal.

### 4.4.2 Race, Ethnicity, and Anglocentrism

From 827 studies, 206 (24.9%) reported on characteristics related to race, ethnicity, and anglocentrism. Of these, 115 (55.8%) related to race, ethnicity, and/or the nationality of participants, while 97 (47.1%) related to language ability and 13 (6.3%) were implied by naming conventions. Notably, race, ethnicity, and nationality were often mixed together and hard to tease apart. For example, Kim et al. [106] reported participants' self-identified racial and ethnic background together, such as "Black or African American" (even though Black people are not necessarily African or American). Additionally, the naming approach, including names but also labels and descriptors, acted as a cue to an anglocentric framing. For example, Ghazali et al. [107] named their agents Mat and Oliver, while Chita-Tegmark et al. [108] named their robots Bob, Jessica, Peter, Katie, and so on. In 15 cases (7.3%) there was insufficient reporting on one or more characteristics, even while others were reported on.

While a diversity of languages, races, ethnicities, and nationalities, and cultural framings were reported on, white English-speakers of anglocentric background were over-represented. 18 studies included white people (8.7%), 63 included English-speakers (30.6%), and 85 (41.3%) included those of other anglocentric backgrounds, such as German-speakers or participants located in Denmark. Overall, 151 studies (73.3%) reported on some combination of white, English-speaking participants with anglocentric backgrounds. For comparison, 124 studies (60.2%) reported on participants who were not white, not necessarily English-speaking, and not of anglocentric backgrounds. Of these, 15 studies (7.3%) reported on non-white participants, 94 (45.6%) reported on participants using a language other than English, and 64 (31.1%) reported on non-Anglo-Saxon backgrounds. No single study reported on all factors: race, ethnicity, language, and cultural background. It is therefore difficult to draw firm conclusions.

### 4.4.3 Age

Of the 827 studies, 787 studies (95.1%) reported the age of recruited participants using one or more of three parameters: mean, standard deviation (SD) and age ranges (min. and max.). Notably, not all studies reported all three parameters. In terms of mean, 391 studies (47.2%) out of 827 reported this parameter. Of these, 134 studies (34.2%) involved youth with mean ages ≤ 24. Regarding SD, 296 (35.7%) out of 827 studies reported this parameter. When describing the age ranges, 349 studies reported the minimum age while 332 reported the maximum age. When describing the minimum age, 315 out of 349 studies (90.2%) recruited youth (aged ≤ 24) and 48 studies (out of 331, 14.5%) included youth within the maximum range of ages. In 332 studies (40.1%), both age ranges were provided; of these, 44 (13.2%) recruited only young participants (aged ≤ 24).

### 4.4.4 Sexuality and Family Configuration

Relatively little was reported on sexuality and/or family configuration (14, 2.8%). Hoffman et al. [94] explicitly reported on "heterosexual" couples, although it is not clear why non-heterosexual couples were excluded, suggesting a normative framing of sexuality. In most cases, sexuality was implied but obscured. For instance, Sung et al. [109] reported on the sex/gender of participants recruited across households, people married and several having at least one child in their care, but without reporting on the sex/gender makeup of the couples or their sexuality Only "married" couples were invited to participate, potentially excluding other valid familial configurations of people who may not have had the legal ability or moral interest in marriage. Similarly, Ostrowski, Breazeal, and Park [110] reported on older adult participants who were living alone, widowed or divorced, or living with a spouse, but did not report on the sex/gender or sexuality of each person or couple. It is not clear whether "widowed" and "divorced" included gay marriages or partnerships. Moreover, generational differences towards sex/gender and sexuality could help explain the proportion of those living alone or reveal important relationships that may not check the typical relationship boxes, e.g., "married" and "divorced."

### 4.4.5 Disability

Only 150 (18.1%) studies reported on participant disability status. Moreover, in 69 of these (46%), no details were provided. Descriptions of study designs indicate the likely exclusion of disabled participants. Several studies assumed that participants were able to watch a video and answer questions concerning its content [111–113]. Others, e.g., [114], tasked participants with observing the behavior of a robot. Such tasks assume that the participants have unimpaired sight, and there were no mentions of how a similar task could be made accessible to a disabled participant interested in the study. Overall, assumption of sight was the most common. Other studies made similar assumptions in relation to hearing [115–117]. Multiple studies, e.g., [118], involved measuring the effects of the robot's voice. This implies that participants would be able to hear and understand what the robot said. Other studies, e.g., [119], involved the robot moving alongside a "walking" participant, which implies that participants would be expected to have a "normal gait" as well as use of their legs for movement.

Thirteen studies specified inclusion criteria that purposely excluded participants with certain disabilities or neurodivergency. For example, the authors of one study [120] stated that participants did not have "known neurological or physical injury that could affect their haptic sensitivity and their physical behaviour" (p. 266). Similarly, those of study [121] reported that "only participants with normal or corrected normal vision could take part in the study" (p. 75). Conversely, 14 studies reported measuring either disability or different types of impairments. However, it is unclear if these were done as part of screening procedures or for demographics. For example, the authors of one study [122] reported that none of the 22 older adult participants had a cognitive impairment, but not if cognitive impairments led to the exclusion of participants.

Only 39 studies directly reported on disability status: participants who were visually impaired [123, 124], autistic [125, 126], hearing impaired [127], mobility impaired [128, 129], cognitively impaired, including dementia [130, 131], medical conditions including cerebral palsy [132, 133] and Parkinson's disease [134], and/or generically described as having disabilities [135, 136]. 18 described participants as healthy [137, 138], able-bodied [139, 140], or typically developing [141]. Definitions for these terms were not provided. Moreover, 12 of these would have benefitted from the inclusion of disabled participants, as the goal was to develop technologies for disabled populations [140, 142, 143]. Also, in 13 studies, we found instances of ableist language. For example, the authors of one study [144] stated that participants had "a typical characteristic of low-functioning autism." Concepts such as high and low functioning autism" (p. 173) have been heavily criticized by advocates as reinforcing stereotypical and medicalised views of individuals that have been repeatedly used to oppress minorities [145]. Similarly, the authors of another study [146] used the term "handicapped" to refer to people who use wheelchairs. The use of this term has also been heavily criticized for reinforcing medical models of disability. Indeed, many have advocated for the discontinuation of its use since the introduction of the International Classification of Functioning in 2001 [147].

### 4.4.6 The Body

A total of 102 (12.3%) studies included information about participants' bodies, and 725 (87.6%) did not. Of those that did, only 43 (42.2%) reported on size, shape, and physical characteristics explicitly; the remaining 59 (57.8%) made implicit assumptions. Most details were linked to the performance of specific actions in the experimental procedures. 55 studies featured implicit assumptions about or specific requirements for participants' bodies, including "features" and capabilities, in inclusion and exclusion criteria. For example, one study [148] assessed the effect of an impolite robot's encouragement on participants' performance of squatting exercises. The authors did not report on specific requirements and/or baseline physical capabilities. Yet, squatting in a "standard fashion" requires participants to have legs, sit and stand repeatedly, and maintain their balance. In contrast, the authors of one study [100] specifically stated that they made sure that potential participants had the ability to wave their hands or perform similar movements as a pre-requisite to being able to initiate or respond to social greetings.

Sixteen studies featured implicit or explicit details on participant size and bodies. For example, one paper [149] describes a study in which the GypsyGyro-18 motion capture suit was used to track the movement of an individual in a robotic workspace. However, to date, most motion capture suits that incorporate Inertial Measurement Units have primarily been used exclusively with participants who have normative bodies, ones that move according to "normalised notions of bodies and movement" [150]. In a similar fashion, some studies [99, 116, 151] that involved the performance and mimicking of a set of gestures associated with particular expressions were designed around normative ideas of what is considered standard body language [42].

Height and handedness were the body factors most frequently explicitly reported. Seventeen studies included information about standard body height and/or efforts to accommodate potential height variations. The authors of one study [152] assessed how the relative height of a robot influences user perceptions about its authority. They explicitly stated that the two heights tested were 188 cm, or the height of the badminton coach that the robotic avatar was representing, and 153 cm, or the average height of a Korean sixth-grade student. In contrast, although they do not report on the robot's height, the authors of [153] explained how their robot featured a touchscreen built to be height-accessible to walking and wheelchair-using participants alike. Finally, 19 studies [154–156] reported on participant handedness, albeit with no mention of ambidexterity; only one study [157] mentioned one ambidextrous participant.

### 4.4.7 Ideology

95 studies (11.5%) reported on matters related to ideology. 728 out of the total 827 studies (88%) did not. Specifically, three studies reported on religion or spirituality, two on politics, 19 on morals and ethics, three on law and policy, 15 on identity, and 18 beliefs and values. In virtually all cases, the reasons for collecting this information were not discussed. While reasons may be implied by the goals of the study or in the research methods, we cannot be sure that our ideological foundations and assumptions are in sync with those of these authors. Thus, we focused on highlighting valid alternatives as well as proceed with caution in reporting these results.

Assumptions about beliefs, value systems, and ideology varied. For instance, van Der Putte et al. [158] asked about "religion or belief" in the context of a health information elicitation robot at a hospital. Yet, it is not clear what relevance religious identity or beliefs have to this task or the data being collected. In contrast, Bartneck et al. [159] elicited perceptions of race-based aggression potential, which was directly connected to the study goal of exploring racialized violence in robots with race cues. Others made assumptions about general values and beliefs. For instance, Powers and Kiesler [160] explored a robot that gives health advice, but relied on body mass index (BMI), which has long been criticized as a flawed measure of obesity, let alone health status, as well as arguably racist and sexist, given its foundations and coordination around the bodies of white men [161]. In another example, Rossi et al. [99] assumed that a bar context involving alcohol would be ethically neutral and lead to results for robots in service contexts generally. Prescriptions of identity also appeared. Cheon and Su [162], for instance, decided to call their HRI research participants "roboticists" without providing a reason. While these results may point to biases on the part of the authors, they also raise opportunities for collaborations with critical scholars and epistemologists.

We also discovered two patterns representing assumptions of universality. The first was in terms of methods (18 studies). Rea, Schneider, and Kanda [148], for instance, asked participants to rate the relative politeness/rudeness of the phrases used in the study, rather than prescribe this characteristic based on their own perspectives. Others, however, made assumptions about the universality of scenarios, such as moral judgments [163], the trolley problem [164], the Desert Survival problem [108], and the Monty Hall problem [165]. Emotional expression and interpretation, which may not be universal across cultures, was also often assumed as generalizable in the use of measures, such as emotional intelligence [166] and robotic expressions of hostility [167]. The second pattern is about the presumed influence of pet ownership (5 studies). Only one paper explicitly mentioned this as the goal of the study [168]. In the rest, no reason was given for why pet ownership was collected [110, 169–171]. The assumption may be that robot ownership is similar to pet ownership, e.g., pets and robots are semi-autonomous dependents over which people have control and rely upon for specific functions in their lives. However, this requires full disclosure and deeper engagement with the reasons underlying this proposed connection.

### 4.4.8 Domain Expertise

Of the 827 studies across 749 papers, only 323 reported on the degree of familiarity and expertise that participants had in relation to computers and robots. In total, 504 (60.9%) studies did not report any information pertaining to the degree of computer literacy or robotics expertise of participants. From the other 156 studies, we were able to discern if participants were computer users (e.g., of smartphones and video games). However, only 72 (46.2%) of these studies explicitly included these details; for the remaining 84 (53.8%), we drew implicitly from other methodological details. For example, in absence of other specific information, when studies mentioned that participants had been recruited through social media, participation involved the completion of online surveys, or experiments had been carried out on digital platforms such as Amazon Mechanical Turk or Prolific, we were able to determine that participants had sufficient familiarity with digital technologies, such as personal computers or smartphones, to be able to access either recruitment adverts or experimental platforms.

Of the 72 studies that reported on technology usage and expertise, 30 focused specifically on computers, whereas 53 included details about other types of technology usage. These groups were not mutually exclusive; several collected this data alongside other details, e.g., interest in video games. Information about "other types" of technology usage were generic or specific. Several only referred to participants' overall usage, self-reported expertise, or familiarity with technology in general. For example, two papers [172, 173] indicated recruitment of participants with "*high technology acceptance*" but without details on the type of technology. Other studies focused on participants' familiarity with broad subsets of technological products, such as video games [171, 174], VR systems [105, 175], smartphones [176, 177], and social media [178, 179]. Some studies covered details about familiarity with specific types of technology, such as the Rviz visualization widget [180], maps in the game Unreal Tournament 2004 [181], and Alternative Augmented Communication (AAC) products [128]. Frequent or expert users were reported as participating more often than novices. In total, 58 studies included participants that were either frequent users (reported usage described as frequent or at least occurring once a week), or moderately familiar/expert users. In contrast, only 14 studies explicitly reported the inclusion of non-users, infrequent users, or novice users (not mutually exclusive).

In 147 studies, the degree of familiarity participants had interacting with robots was reported. 109 studies focused on robots as a general category of artifacts, without distinguishing between different types of robots. Alternatively, 47 studies covered specific sub-categories of robots such as social robots [182, 183], drones and aerial robots [184, 185], NAO robots [186, 187], or Pepper robots [188, 189]. Moreover, some authors reported on participants' previous interaction (or lack thereof) with the specific robot developed used in the study. This was generally indicated by statements that participants "had never interacted with our robots before" [190]. Finally, 35 studies explicitly mentioned participants who had specific expertise as roboticists alongside non-experts, e.g., [191, 192], and often as the primary target group, e.g., [193–196]. Only six studies reported purposefully excluding those with robotics expertise [197–201].

Measurements of familiarity and expertise were varied, making comparison across studies challenging. For example, Likert scales varied from 7-point, e.g., [202], e.g., 5-point [203], and 3-point, e.g., [204]. Others [205, 206] used classifications, such as experts and novice users. Still others [177, 207] simply reported the presence or absence or previous interactions between participants and robots. With this in mind, we identified 69 studies reporting that at least some participants were already familiar with robots and 97 where at least some were not. A number of these overlap, as it was not uncommon for researchers to purposefully include participants who had varying degrees of familiarity with robots [208].

In total, 71 studies reported collecting information on participants' current field of study, their background, and/or their experience with programming. We found that 50 studies specifically included at least some engineering and/or computer science students [209, 210], researchers [211, 212], and/or individuals more generically described as having a technical background [213, 214]. Only 15 studies included participants who did not have an engineering or computer science background. In five cases, this was part of diverse sampling strategies [162, 213, 215–217]. In 15 studies, participants disclosed experience with programming, but measurements varied: Likert scales [217, 218], nominal categories [191, 219], or no details [220].

### 4.5 Influence of Length of Paper

Out of 749 papers (827 studies), 423 papers (438 studies) were classified as short and 326 papers (389 studies) were classified as long. Table 4 shows the total number of studies by length of paper against the WEIRD framework. Table 5 shows the same for each diversity factor as well as by the detailed coding classification for each factor. Short papers were less likely to have details related to participant WEIRDness. Only 23.7% of these included either the nationality of participants or the location of the study. In comparison, 39.8% of long papers included this information. Similarly, 38% of long versus 27.9% of the short papers reported information on education level. Diversity factors were also more likely to be reported in long papers. For certain factors, such as the body (12.6% of short vs. 12.1% of long) and disability (18.7% of short vs. 17.5% of long), this discrepancy was relatively small. However, for all other factors the differences were much greater. The greatest discrepancies were for domain expertise (2.3% of short papers vs. 82.3% of long), and age (10% of short vs. 78.4% of long). This suggests that when there is room certain factors are more likely to be reported on than others, and these appear to be the most common and least sensitive factors.

**Table 4** Counts of studies in short and long papers for WEIRD variables

|  | Western | Education level | Industrialization | Rich status | Democratic status |
| --- | --- | --- | --- | --- | --- |
| Short (reported/unreported) | 104/334 | 122/316 | 104/334 | 104/334 | 104/334 |
| Long (reported/unreported) | 155/234 | 148/241 | 155/234 | 155/234 | 155/234 |

### 4.6 Archetypes: WEIRD and Diverse

Who is the most represented participant archetype in the corpus? Our results paint an emerging picture of a WEIRD and "uncanny" character. Given the state of reporting, we must take this archetype with a grain of salt. Nevertheless, as Tables 1 and 2 reveal in summarized form, the most common participant is likely to be from the West, especially the US, and located in an industrialized, rich nation. They are likely to be anglocentric in identity and origin, especially English-speaking and white-identifying. They are apt to have a high level of political freedom and be highly educated, as well as knowledgeable when it comes to technology and/or robots. They are less likely to have a disability because researchers implicitly or explicitly create barriers to inclusion based on body, size, and/or disability. We may know something about their morals, beliefs, and identity, but little about their religion, politics, and standing on legal matters. They may identify as a man or a woman, or be sexed as male or female, but we may not know if they identify in a more gender diverse way, and we will not know their sexuality. They may not be very young, but they will likely be a young adult and certainly not an older one.

Who is the least represented participant archetype? It is difficult to draw firm conclusions because of limitations in reporting. We can reverse what is known, Venn diagram style. This indicates that we are less likely to find people located in the East who are not white, unless they are from Japan and Korea. We are not likely to find uneducated people or even people with low education. We are less likely to recruit from poorer nations with low democratic empowerment and reduce industrial output. We are less likely to include people of size and disabled people, unless the study focused on disabled people. We are less likely to involve non-experts and extreme experts, paradoxically. We are less likely to recruit gender-diverse people of various sexualities. We may recruit people who know various languages, but none will reach the frequency of English speaker representation. We are not likely to find older adults. However, we must take these results with a grain of salt. Put simply, *lack of reporting does not equal lack of inclusion*. We discuss the implications and possible ways forward next.

## 5 Discussion

We undertook this work to assess the human side of the human–robot interaction equation in research. Driven by critical scholarship within and beyond engineering and computer science, we asked: Are HRI participants WEIRD? However, we realized that this question alone was insufficient to assess how "weird" HRI research populations might be. We thus went a step further: Are HRI participant samples strange in other ways? Are they diverse? Our systematic review reveals that HRI participants are indeed WEIRD and may also not be as diverse as expected. Even so, we must consider these results in light of another major finding of this work: the state of reporting in HRI. We now turn to discussing each of these findings and potential ways forward, as well as limitations of our own work and trajectories for future work.

### 5.1 WEIRD and Diverse Patterns in HRI Participants

HRI participants are on the majority WEIRD or EIRD, located in or of nationalities that are primarily Western, and especially American, with the notable exceptions of Japan and Korea (as the EIRD nations). Given the global powerhouse that is the US, this is not unexpected. Moreover, Japan and to some extent Korea have long been heralded as technology-forward nations. Japan, in particular, is globally recognized for its contributions to robotics [221]. We thus might not be surprised to find over-representation in HRI samples from these countries. A related pattern on the diversity factors side is apparent anglocentrism. A rather "weird" aspect of WEIRD research is the lack of reporting on race, ethnicity, and cultural identity or background. In particular, "the West" seems to be a code for people of certain characteristics, although this is not explicitly represented in the WEIRD framework. Recent critical race scholarship would argue that these characteristics are whiteness, English ability, if not nativity, and Anglo-Saxon heritage. Our sample indicates that this could be true for HRI research. At the same time, we should be mindful of other possibilities. Some factors may be sensitive, such as matters of disability or the

**Table 5** Coding classifications used for diversity factors, with counts of studies for short and long papers

| Coding classification for the diversity framework | Counts for short papers* | Counts for long papers* | Total counts reported/unreported |
|---|---|---|---|
| *The body* | | | SP: 55/383 LP: 47/342 |
| Implicit/explicit | 35/19 | 24/24 | |
|   Dominant hand | 8 | 11 | |
|   Body weight | 1 | 1 | |
|   Height | 13 | 4 | |
|   Movement | 29 | 26 | |
|   Normative body language | 7 | 9 | |
| *Ideology* | | | SP: 10/428 LP: 85/304 |
| Religion | 1 | 2 | |
| Politics | 1 | 1 | |
| Morals | 1 | 18 | |
| Law | 0 | 3 | |
| Identity | 0 | 16 | |
| Beliefs | 0 | 18 | |
| Compensation | 4 | 19 | |
| Universals | 3 | 20 | |
| *Domain expertise* | | | SP: 3/435 LP: 320/69 |
| Implicit/explicit | 1/0 | 83/72 | |
|   Computer familiarity | 1 | 155 | |
|   Robot familiarity | 1 | 146 | |
|   Students in CS/Eng. | 0 | 71 | |
|   Experts who create/research | 1 | 34 | |
|   Experts purposefully excluded | 0 | 6 | |
| *Race and ethnicity* | | | SP: 25/413 LP: 181/208 |
| White/non-white | 1/1 | 17/14 | |
| English/not English | 3/13 | 60/81 | |
| Anglo-centric/not anglocentric | 19/3 | 66/61 | |
| Language | 8 | 14 | |
| Race/ethnicity | 2 | 89 | |
| Nationality | 12 | 18 | |
| Name | 3 | 83 | |
| No details | 1 | 10 | |
| *Sex and gender* | | | SP: 20/418 LP: 136/253 |
| Gender | | | |
|  Expansive | | | |
|   Inclusion | 4 | 13 | |
|   Privacy | 4 | 31 | |
|  Binary | | | |
|   Balance | 1 | 31 | |
|   Women only | 8 | 53 | |
|   Male only | 0 | 13 | |
|   Disproportionate Gendering | 1 | 4 | |

**Table 5** (continued)

| Coding classification for the diversity framework | Counts for short papers* | Counts for long papers* | Total counts reported/unreported |
| --- | --- | --- | --- |
| Unconscious | 2 | 2 | |
| (Cis)sexist | 0 | 6 | |
| *Sexuality and family configuration* | | | SP: 4/434 LP: 10/381 |
| Normative | | | |
| Heteronormativity | 0 | 2 | |
| Obscured | 4 | 8 | |
| *Disability* | | | SP: 82/356 LP: 68/327 |
| Assumptions causing exclusion | 36 | 33 | |
| Purposeful exclusion | 6 | 7 | |
| Features of non-neurotypical and disabled participants | 26 | 13 | |
| Healthy participants | 14 | 4 | |
| Ableist language or implications | 8 | 5 | |
| Unclear if screens or assess | 3 | 11 | |
| Should not exclude disabled participants without fair and unambiguous reasoning | 9 | 3 | |
| *Age* | | | SP: 44/394 LP: 305/84 |
| Youth | 16 | 118 | |
| Emerging adulthood | 36 | 221 | |
| Older adults | 5 | 14 | |

*Counts for individual reported factors in the diversity framework are not mutually exclusive

body, and researchers may be uncertain whether and how to ask. As we will discuss, institutions could also place limits and barriers on demographics data collection that mask a true desire on behalf of researchers to capture this data for the purpose of reporting on representation. Moreover, we do not have a formal way by which to capture and represent most of this information, which we discuss next. Yet, there is a shift occurring, with recent work tackling race, ethnicity, and identity in HRI spaces head-on. But we cannot just work on "robots and race;" we also need to capture the "race and humans" element in our samples.

HRI samples are also "uncanny" in other ways related to factors of diversity. Robots are physical, for the most part, and robots that interact with humans often do so through physical means (but not always, as is the case with conversational robots). Yet, participant embodiment was vastly under-considered. This is despite a surge of research on factors of the body, especially approach distance [222], medical robots that lift people [223, 224], and recognition that the relative size of the robot can instill comfort or discomfort [152]. Moreover, how disability plays out and what characteristics of the body relate to disability were almost never considered. We also found evidence of exclusion based on researcher assumptions of ability and "health," as well as explicit exclusion based on bodily features and/or disability. We urge our fellow researchers to include people of all configurations, unless they have a good reason not to. This may mean recognizing that the research is disabling in some way and correcting it. If certain bodies or embodiments need to be excluded, clear and fair reasoning should be provided. We should never exclude based on convenience.

Virtually all researchers have taken an ideologically neutral approach, relying on an assumed foundation of beliefs, attitudes, opinions, and values. This tended to occur even when researchers were conducting research on morality and ethics. Yet, the relationship between ideology and beliefs and the research at hand may be difficult to determine. At the very least, we encourage researchers who study robots and ethics, law, morals, beliefs, religion, spirituality, and other ideological topics to capture the relevant demographics and attitudes, incorporating these in their analyses, and reporting on them faithfully. At the same time, we should consider the ethics of asking about personal ethics. We should avoid forced disclosures, not only for the comfort and safety of participants, but also to ensure data quality. We refer to other work [16, 225, 226] on how to navigate this sensitive aspect of reporting.

While WEIRD research has highlighted the problem of relying on undergraduate populations, this issue has been under-acknowledged in HRI research. Yet, as our critical review shows, HRI samples have been primarily made up of people who knew computers well, were students in computer science or engineering, and were familiar with robots. Familiarity can bias results, and we must take heed not to over-generalize our results, given the over-representation of "those in the know" as participants. Additionally, HRI samples tend to be young. Given the average age of undergraduate students in most nations, this is to be expected. Nevertheless, we should aim to capture a full range of human experience, across age groups, and without making assumptions of interest or ability based on age, as some in our corpus have done.

When it comes to sex, gender, sexuality, and family configuration, the expected patterns exist. Most research has relied on limited frameworks of sex and gender. While this is changing, the conflation of sex and gender and reliance on the gender binary prevails. We found almost no reporting of non-binary and transgender people, or gender diversity. This does not mean that diverse people were not included, as it was also difficult to access how data was collected. Moreover, many relied on an "other" category, which collapses diversity and implicitly "others" people, i.e., acts as a cue that the person is atypical [50]. We do not know if it was a recruitment problem or a measurement problem, i.e., gender-diverse options were not offered. HRI researchers can follow recent shifts on asking about and reporting on sex/gender [11, 47, 49, 50]. Finally, we discovered several meta-level patterns resulting from social norms and habits in HRI or research generally that should be highlighted and challenged. Many researchers, operating from a gender binary perspective, only reported female or women counts and/or percentages. The implication is that "the rest" are male or men. This may be a matter of social norms in research reporting, arising from a legacy of women being excluded as participants and, which was normalized over time, with the goal of highlighting the recruitment of women. Even so, our analyses show that researchers did this even when there were more women or girls recruited than men or boys. Moreover, there were roughly even numbers of men and women participants overall. We raise this question for the community: Why continue? For breadth and accuracy, we recommend reporting on whether sex and/or gender was captured, and then providing the counts and/or percentages for each sex (as there are a range of intersexes) and gender, making explicit note of whether diverse gender identities were considered. This should be reported for the sake of the research goals as well as for transparency in representation, regardless of whether the data is used for main analyses.

The apparent lack of diversity has special implications for HRI research. Robots are often humanlike and social, and this matters. We draw from the Computers are Social Actors (CASA) paradigm [227, 228], which is backed by a wealth of research over the last couple of decades [229, 230]. In short, the research indicates that we tend to ascribe and react to human-like computer agents as if they are human, often without realizing it, and sometimes even when we do. (This may in fact be an argument in favour of not worrying too much about the overrepresentation of people familiar with robots in HRI samples–they are not necessarily immune to this phenomenon.) Robots are also expensive and typically built for WEIRD nations or the nations in which the builders are located. This can have implications beyond cost, including language support, local tech support, and so on. Moreover, the very notion of "robot" may be "WEIRD" and certainly has "WEIRD" roots. Nevertheless, robot-adjacent concepts may exist, and the robot concept itself can be adapted elsewhere. This raises important questions about research inclusion, not only for participants but also the researchers themselves.

What can we do? In their CHI paper, Linxen et al. [3] provide several ideas that may be appropriate for HRI venues, too: diversifying authorship; fostering the use of online research; developing methods for studying geographically-diverse samples; appreciating replications and extensions of findings; reporting and tracking the international breadth of participant samples; and identifying constraints on generalizability. We echo these suggestions with some caveats and additions. Online research, for instance, was reported in 77 studies (9.5%), and we expect this proportion to increase as a result of shifting attitudes towards research practice following the global COVID-19 pandemic [231]. The challenge will be how to incorporate physical robots into virtual or hybrid research contexts. Other challenges remain. HRI research has not been widely conducted outside of WEIRD and EIRD nations. We imagine two opportunities here. First, WEIRD and EIRD researchers can make a concerted effort to bring on researchers, labs, companies, and institutions as collaborators. Second, we may seek to learn from non-WEIRD and non-EIRD researchers and participants about what robots are or can be. Wealth of all kinds can be shared … including intellectual, artistic, and phenomenological wealth. We can also take up posts as outreach officers, such as for ACM and IEEE regional chapters. We also add a call for reflexivity and stricter reporting standards. Perhaps this is a matter of under-reporting, which could shift the results either way. We turn to this topic and propose a solution next.

## 5.2 A Matter of Reporting?

We have a reporting problem in HRI research. Much of our analyses, and therefore our results, were limited by insufficient reporting. This played out in a variety of ways. Some researchers simply did not report any information about participants. In some cases, there was no mention of participants in the paper, and we had to make a guess based on what was implied by other features of the research, such as the system design, data analyses, and results. Others reported some information but not all (e.g., 86% were students, but who were the rest?). Others reported information in non-standard ways or ways that cannot be used in meta-analyses (e.g., median ages). Some information was obscure due to lack of detail (e.g., nationality or location? Does "other" mean another gender identity or that someone preferred not to say, or something else?). There was also unclear or implicit reporting, (e.g., "roughly" 100 participants). If this state of affairs continues, then we will not be able to determine the extent of the underlying problems, or lack thereof, when it comes to representation and inclusion.

All of these issues are easy to fall prey to … but potentially easy to resolve, at least in theory. Indeed, the greater scientific community, notably headed by Nature group, have recently made strides towards improving reporting by providing templates.[15] Other fields of study, in particular the medical and health fields, have long recognized the need for standard reporting to evaluate the relative degree of consensus on a certain intervention. PICO (Population, Intervention, Comparison, Outcomes) [232], PICOS (PICO plus Study) [233], and SPIDER (Sample, Phenomenon of Interest, Design, Evaluation, Research type) [234] are long-standing, widely used templates in these domains. Nevertheless, they are disciplinary and high-level. Moreover, HRI papers are as likely to be short papers as long papers. There may simply not be enough space to report all details. Indeed, the reported counts for the short and long papers suggest that researchers may have been forced to cut details due to space limitations. Finally, we acknowledge that there may be institutional and structural barriers to capturing and reporting participant details. For example, ethics boards may request or require a limit to the number and kind of demographics questions asked. Such barriers may not be resolvable, but should be reported as an explanation, e.g., "The ethics board did not allow us to capture demographics factors that were not directly tied to our research questions and hypotheses."

Keeping in mind the particularities of the HRI conference, we offer our recommendation. First, consider adopting an existing template. SPIDER may be especially reflective of most HRI research. Adapting a template or developing a new one will take time and community engagement. Future work should involve workshops and other forms of engagement as well as testing templates out. Ideally, the HRI conference will develop a standard template and provide it in the template for papers. This may be especially important for short papers, which can be as few as two pages. The first page could use a template like SPIDER and the second page could be open-ended, based on the characteristics of the reported research. Second, we recommend using the WEIRD and diversity frameworks as a checklist and format for reporting. We offer the following structure for writing up results based on the clusters and intersections among WEIRD and diversity factors, with sensitive or case-dependent factors in square brackets:

> Age, [sex], gender, [sexuality], [family configuration], race, ethnicity, nationality, location, education, computer-oriented education, [ideology], disability status, [body factors]

Regardless, we need to report whether our recruitment measures were successful, as well as when they were not. We need to report upon failure to recruit, rather than leave it out and allow the reader to assume. For example, one study [192] reported on failure to recruit ideal participants: "Unfortunately, we were unable to recruit a guide dog user" (p. 107). This is clear and to the point. We urge other researchers to do the same. In a similar fashion, we acknowledge that it might not be possible, or even appropriate, to collect information concerning all diversity factors of participants. Potentially sensitive and/or uncomfortable questions on sex, sexuality, race/ethnicity, and ideology (some of which have legal ramifications in certain nations) might justifiably raise ethical concerns, especially when they are not directly linked to the research question of a study. However, to increase clarity and transparency in the reporting, we believe that is important for researchers to specifically state when they choose not to collect information about diversity factors and, where possible, the reason behind their choice.

## 5.3 Limitations and Future Work

We did not cover all HRI venues or venues containing HRI work, given the sheer volume of papers that would result. Future work can assess these venues. We did not extract all pertinent information, such as where the conference was held each year, where the authors were from, and demographics on the institutions to which each author belonged. Additionally, we did not conduct intersectional analyses due to lack of scope and space. Future work can identify and map out how a greater breadth of intersecting characteristics are, or are not, represented in the literature. Finally, given the state of reporting, we were not able to draft up statistics and/or clear findings on all characteristics and intersections. We expect that more rigorous reporting going forward will allow for this. To this end, we can test the diversity template above for reporting on sample characteristics. Future work can refine this template for optimal reporting.

---

[15] https://www.nature.com/documents/nr-reporting-summary-flat.pdf.

## 6 Conclusion

Robots and people are at the heart of HRI research. Yet, the samples making up an important and influential portion of HRI research appear to be not only WEIRD but lacking in diversity, as far as reporting allows us to gather. We offer our results and suggestions with humility and a keen desire to improve what is already an excellent corpus of work. We, the authors of this work who also have work in this corpus, are no exception. We will continue to be reflexive and change our practice. Human knowledge (and robot knowledge) is of us, for us, and by us all. We should not be rattled or dismayed, but instead accept that reality is messy. Frameworks like WEIRD and our diversity factors can help us make sense of it and map it out in our research. We can start by acknowledging the "who" of HRI research with greater rigour and transparency. Now that we know the state of affairs and have an idea of what to do about it, we can start taking steps as a community of practice towards a more representative and diverse future.

**Acknowledgements** This work was supported in part by Japan Society for the Promotion of Science (JSPS) Grants-in-Aid for Scientific Research (#21K18005 Early Career Scientists and #PE20728 Postdoctoral Fellowship) and funding from the Canada 150 Research Chairs Program.

## Declarations

**Conflict of interest** The authors declare that they have no conflicts of interest.

**Publisher's Note** Springer Nature remains neutral with regard to jurisdictional claims in published maps and institutional affiliations.